\documentclass[doublespacing]{elsart}
\usepackage{epsfig}

\begin{document}

\center

\maketitle
\begin{frontmatter}
\title{Neutrino Detection With CLEAN}

\thanks[corr]{Corresponding author D.N. McKinsey. Email: 
daniel.mckinsey@yale.edu}
\author[yu]{D.\,N. McKinsey} and
\author[nist]{K.\,J. Coakley} 
\address[yu]{Yale University, New Haven, CT 
06511, USA}
\address[nist]{National Institute of Standards and Technology,
Boulder, CO 80305, USA}

\begin{abstract}
    
This article describes CLEAN, an approach to the detection of
low-energy solar neutrinos and neutrinos released from supernovae. 
The CLEAN concept is based on the detection of elastic scattering
events (neutrino-electron scattering and neutrino-nuclear scattering)
in liquified noble gases such as liquid helium, liquid neon, and
liquid xenon, all of which scintillate brightly in the ultraviolet. 
Key to the CLEAN technique is the use of a thin film of
wavelength-shifting fluor to convert the ultraviolet scintillation
light to the visible.  This allows the same liquid to be used as both
a passive shielding medium and an active self-shielding detector,
allowing lower intrinsic radioactive backgrounds at low energies.

Liquid neon is a particularly promising medium for CLEAN. Because
liquid neon has a high scintillation yield, has no long-lived
radioactive isotopes, and can be easily purified by use of cold traps, it
is an ideal medium for the detection of rare nuclear events.  In
addition, neon is inexpensive, dense, and transparent to its own
scintillation light, making it practical for use in a large
self-shielding apparatus.  If liquid neon is used in CLEAN, the center
of the full-sized detector would be a stainless steel tank holding
approximately 135 metric tons of liquid neon.  Inside the tank and
suspended in the liquid neon would be several thousand
photomultipliers.

Monte Carlo simulations of gamma ray backgrounds have been performed
assuming liquid neon as both shielding and detection medium.  Gamma
ray events occur with high probability in the outer parts of the
detector.  In contrast, neutrino scattering events occur uniformly
throughout the detector.  We discriminate background gamma ray events
from events of interest based on a spatial Maximum Likelihood method
estimate of event location.  Background estimates for CLEAN are
presented, as well as an evaluation of the sensitivity of the detector
for $p-p$ neutrinos.  Given these simulations, the physics potential of
the CLEAN approach is evaluated.

\end{abstract}

\end{frontmatter}



\section{CLEAN}

Just as new physics can be learned by building accelerators with high
collision energies, new physics can also be learned by building
underground detectors with high sensitivity to rare events, such as
neutrino interactions, scattering of dark matter particles, and double
beta decay.

This latter field is now the source of great excitement, as the recent
results of the Super-Kamiokande, SNO, and KamLAND neutrino detectors
have proven that neutrinos have mass\cite{Fuk98,SNO,Egu03}.  This is
the first substantial change in the Standard Model of particle physics
in the last 20 years.  In addition, new highly sensitive dark matter
detectors, such as CDMS, EDELWEISS, and ZEPLIN, have driven down the
limits on dark matter scattering cross-sections, bringing us
significantly closer to theoretical predictions for supersymmetric
dark matter, and therefore closer to testing one of the best
explanations for the missing matter of the
universe\cite{CDMS,Edelweiss,ZEPLIN1}.  In our quest to learn more
about neutrinos and other weakly interacting particles, it would be
extremely valuable to have a better detector technology than is
currently available.  Ideally, this detector technology would
simultaneously provide low radioactive backgrounds, a low energy
threshold, and large detector mass at reasonable cost.

\subsection{Experimental design}

\begin{figure}[tbhp]
    \begin{center}
	\epsfig{file=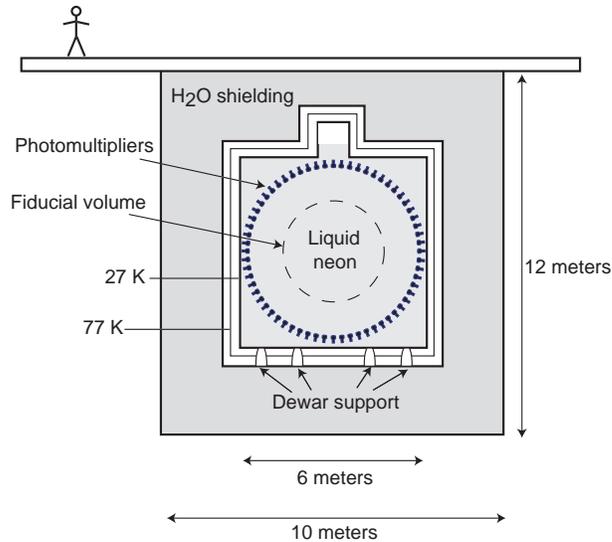, width=8cm}
\caption{A conceptual sketch of 
the full-sized CLEAN apparatus.}
\label{fig:CLEAN_apparatus}
\end{center}
\end{figure}

The following is a description of a scheme that meets all of these
requirements, in which liquid neon is used as a detection medium.  It
is called CLEAN, standing for \underline{C}ryogenic \underline{L}ow
\underline{E}nergy \underline{A}strophysics with \underline{N}oble
gases. 
In a CLEAN detector, neutrino-electron scattering events: $$
\rm \nu +
e^{-} \rightarrow \nu + e^{-}$$
and neutrino-nucleus scattering events:
 $$ \rm \nu + Ne \rightarrow \nu + Ne$$
would be detected using liquid neon as a scintillator.
\footnote{While
the CLEAN concept might be used with liquid helium or liquid xenon
instead, this paper will concentrate on the liquid neon version. 
Liquid neon is significantly denser than liquid helium, while being
easier to purify than liquid xenon, making it (in our view) especially
promising for detection of rare events.} 
The CLEAN concept was first
proposed several years ago\cite{McK2000}, and in the years following
the basic approach has undergone some revision as new ideas have
surfaced.  Our current concept of the final CLEAN detector is shown in Figure
\ref{fig:CLEAN_apparatus}.

In this design, the central detector consists of a stainless steel
tank filled with liquid neon.  Supported by a spherical geodesic
structure would be about two thousand photomultipliers facing the
center of the tank.  Each photomultiplier window would be coated with
a thin layer of organic fluor material, which would convert the
extreme-ultraviolet (80 nm) neon scintillation light to the visible. 
Only the central ``fiducial volume'' of the neon would be used for
detecting neutrinos; the outer ``shielding volume'' would prevent
gamma rays from the photomultipliers and tank from reaching the
fiducial volume.  Around the assembly would be a tank of water 
serving as gamma-ray shielding, neutron shielding, and muon veto.

Liquid neon has several characteristics that make it ideal for the
detection of rare events:

\begin{itemize}
	
	\item Like all of the liquified noble gases, liquid neon scintillates 
	brightly in the hard UV. This translates into a large signal for 
	ionizing radiation events (15,000 photons per MeV)\cite{Mic02}. 
	
	\item Like other noble liquids, liquid neon does not absorb its
	own scintillation light.  The scintillation light results from the
	decay of neon excimers and is of lower energy than that needed to
	excite the ground-state neon atom.  Scintillation light can
	therefore be extracted from the very large detectors that are
	necessary for the efficient detection of rare events.
	
	\item Neon has no long-lived isotopes and therefore has no
	inherent radioactivity.  This stands in contrast to argon,
	krypton, and xenon, which contain natural radioactivity from the
	isotopes $\rm^{39}Ar$, $\rm^{85}Kr$, and $\rm^{136}Xe$.  This also
	sets liquid neon apart from organic scintillator, which inevitably
	contains beta radiation from the decay of $\rm ^{14}C$.
	
	\item Neon has very low binding energies to a variety of surfaces. 
	This allows neon to be very effectively purified of radioactive 
	contaminants by use of cryogenic traps. Liquid helium at superfluid 
	temperatures presents the only superior opportunity for cleaning 
	contaminants\cite{Ada99}. 
	
	\item Liquid neon is dense ($\rm \rho$ = 1.2 $\rm g \cdot cm^{-3}$) and
	therefore uses space efficiently as a self-shielding medium. 
	Liquid helium, in contrast, has a maximum density of $\rm \rho$ =
	0.145 $\rm g \cdot cm^{-3}$, and therefore requires more or different
	shielding.  Liquid xenon has a density of $\rm \rho$ = 3.06 $\rm g 
	\cdot cm^{-3}$, and is even better at self-shielding, but at a greater
	cost and with likely greater contamination by impurities.
	
\end{itemize}

Though this approach requires a cryogenic apparatus (the boiling
temperature of neon is 27 K), this would not make the experiment
overly complex or costly.  For example, a commercial Gifford-McMahon
cooler operating at 27 K with a cooling power of 75 W costs only
\$34,000.  With a total estimated heat load of 200 W in the full-sized
CLEAN, the apparatus described above would derive little of its cost
from its cooling requirements.  The tank could be supported from below
with stainless steel or titanium tubes, as is done commonly with large
liquid helium-cooled accelerator magnets.  Also, such a large
cryogenic detector would not constitute a substantially new technical
challenge; for example, the ICARUS collaboration has successfully
built and tested a time projection chamber filled with 600 metric tons
of liquid argon\cite{Arn03}.  Also, the Fermilab 15-foot Bubble
Chamber, filled with a neon-hydrogen mixture (61.7 \% atomic neon,
38.3 \% atomic hydrogen), had a 23-ton mass when filled, and was
successfully used for neutrino-electron scattering
experiments\cite{Bak89}.

\subsection{Scintillations in liquid neon}

A relatively clear model of scintillations in liquid neon can be
elucidated from the numerous experimental characterizations of
charged-particle-induced scintillation in condensed noble
gases\cite{Stockton72,Packard70,Surko70,Roberts73,Kubota79,Hitachi83,McK03a,McK03b}.
 When an energetic charged particle passes through the liquid,
numerous ion-electron pairs and excited atoms are created.  The ions
immediately attract surrounding ground state atoms and form ion
clusters.  When the ion clusters recombine with electrons, excited
diatomic molecules are created.  Similarly, the excited atoms react
with surrounding ground state atoms, also forming excited diatomic
molecules.  Fluorescence in condensed noble gases is observed to be
almost entirely composed of a wide continuum of EUV light, emitted
when these excited diatomic molecules decay to the monoatomic ground
state.  The energy of emission is less than the difference in energies
between the ground state (two separated atoms) and the first atomic
excited state for any given noble gas.  The scintillation target is
thus transparent to its own scintillation light, and a detector based
on a condensed noble gas can be built to essentially arbitrary size
without signal loss from reabsorption.  The liquid will scatter its
own scintillation light (a recent theoretical estimate predicts a
Rayleigh scattering length of 60 cm\cite{Sei02} for liquid neon), but
this light will eventually reach the walls of the detector and can be
detected.

Recent studies of the scintillation properties of liquid neon have
found it to be an efficient scintillation medium\cite{Mic02}.  These
measurements, when scaled to recent measurements of EUV scintillation
yield in liquid helium, show that roughly 10,000 prompt EUV photons
are emitted per MeV of electron energy.  The wavelength of these
photons was measured by Packard \textit{et al.}, who found that the
electron-excited emission spectrum of liquid neon peaks at 80
nm\cite{Packard70}.  Liquid neon should also have an intense
afterpulsing component due to the extreme ultraviolet radiation of
triplet molecules.  In liquid neon, the ground triplet molecular
lifetime has been measured to be 2.9 $\rm \mu
s$\cite{Mic02,Suemoto79}.  The intensity of this afterpulsing is
approximately 50 \% of the prompt signal, and can be added into the
prompt signal to improve resolution of energy and position.

The liquid neon scintillation light cannot pass through standard
window materials.  Fortunately, recent work towards detection of
ultracold neutrons trapped in liquid helium\cite{Doyle94} has resulted
in the characterization of efficient wavelength shifting fluors that
convert EUV light into blue visible light\cite{McK97}.  This blue
light is well matched to the peak sensitivity of available
photomultiplier tubes.  Tetraphenyl butadiene (TPB) is the fluor of
choice, having a (prompt, $\rm <$ 20 ns) photon-to-photon conversion
efficiency from the EUV to the blue of roughly 100
\%\cite{McK03b,McK97}.  The prompt scintillation component from the
combined liquid helium-waveshifter system has been measured to have a
width of 20 ns, allowing the use of coincidence techniques to reduce
background\cite{McK03a}.  In addition, as in organic scintillators,
pulse-shape analysis could likely be used to discriminate light
ionizers such as electrons from heavy ionizers such as alpha particles
or nuclear recoils.  Pulse shape analysis has proven to be effective
in liquid xenon\cite{Aki02,Ber01} and liquid helium\cite{McK03a}, and
we expect it to be effective in liquid neon as well.  Combining the
prompt and delayed components, one expects about 15,000 photons per
MeV of electron energy.  Given a waveshifting efficiency of 100 \%, a
photomultiplier covering fraction of 75 \%, and a bialkali
photocathode quantum efficiency of 15 \%, a total photoelectron yield
of about 1690 per MeV could be achieved.  With this photoelectron
yield, the energy of a 100 keV neutrino-electron scattering event
could be measured with an average of 169 photoelectrons, with an
associated 1-sigma random uncertainty of 7.7 \%.

\subsection{External backgrounds}

We estimate the locations of ionizing radiation events in CLEAN by
analyzing the pattern of photomultiplier hits \cite{Coa02}.  The
simulation consists of three steps.  First, the external backgrounds
(gamma and x-rays emitted from the photomultipliers) are simulated,
and the locations of energy deposition for each gamma ray are
recorded.  Second, the trajectory of scintillation photons are
simulated as they travel through the liquid,assuming a Rayleigh
scattering length of 60 cm for liquid neon.  When they enter the
wavelength shifter, the photons are re-emitted isotropically (both
toward and away from the closest photomultiplier), and the photons can
be detected when they reach a photomultiplier (assuming a quantum
efficiency of 15 \%).  The photomultiplier hit pattern due to each of
these gamma rays is recorded.  It is assumed that each photoelectron
is recorded, so that each PMT can count multiple hits.  Third, the
event locations are estimated using a Maximum Likelihood method,
yielding an estimated radius $r_{est}$ for the center-of-mass of
energy deposition.  The initial guess for the event location in the
Maximum Likelihood method is the centroid of the multiple PMT
photoelectron distribution.

We define an inner spherical fiducial volume $r \le R p^{1/3}$ that 
occupies a fraction $p$ of the total spherical detection volume defined 
by $r \le R$. ÊIn CLEAN the probability that an external background 
event occurs within the fiducial volume is low since most external 
gamma and x-rays do not penetrate into the inner fiducial volume. 
ÌáIn contrast, events of interest occur uniformly throughout the entire 
detection volume. ÊThus, we can confidently discriminate external 
background events from events of interest that occur within a fiducial 
volume if we can accurately estimate the radial location of an event. 
We assign events into one of two classes. ÊThe first class consists 
of events that occur within the fiducial volume. ÊThe second class 
consists of events that occur outside the fiducial volume. ÊThe rate 
at which we assign external background events to the first class is 
the background rate for the CLEAN experiment. Following \cite{Coa02}, 
we implement a classification rule based on the statistical distribution 
of $r_{est}$ determined from events that occur uniformly throughout the 
entire detection volume. 
In a Maximum Likelihood approach, 
we estimate the $p$th quantile of the radial 
location estimates computed from the calibration data as a function of 
the number of detected photons. Events for which $ r_{est} $ is less 
than this quantile are assigned to the first class. ÊIn \cite{Coa02}, 
we also corrected $r_{est}$ using a polynomial calibration model 
to account for the effect of scattering and wavelength shifting. 

To speed up our Monte Carlo simulation code, we modify the Maximum
Likelihood approach as follows.  We estimate the radial location of
events in the calibration data set using a polynomial prediction model
based on the centroid of the hit pattern (\cite{Coa02}).  If this
estimate exceeds $\beta R$ (where $\beta =$ 0.8 for the cases studied)
we assign the event to the second class.  Define the fraction of
events for which the Centroid method estimates exceeds $\beta R$ to be
$q$.  For each event in the other $1-q$ fraction of the calibration
data, we compute a Maximum Likelihood estimate $r_{est}$, and the
$p/(1-q)$th quantile of these estimates.  Any event for which
$r_{est}$ is less than this quantile is assigned to the first class.

\begin{figure}[tbhp]
    \begin{center}
\epsfig{file=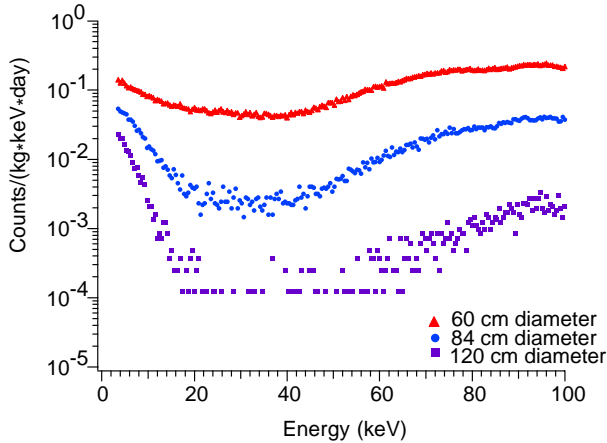, width=8cm} 
\caption{External background
rates in several sizes of future CLEAN detectors.  Shown are estimated
background rates for detectors of 60, 84, and 120 cm diameter,
corresponding to neon masses of roughly 120, 350, and 1000 kg. 
Results correspond to a fiducial volume of $r$ $\leq$ 12.6 cm, with a
fiducial mass of 10 kg.}
\label{fig:background}
\end{center}
\end{figure}

\begin{figure}[tbhp]
\begin{center}
\epsfig{file=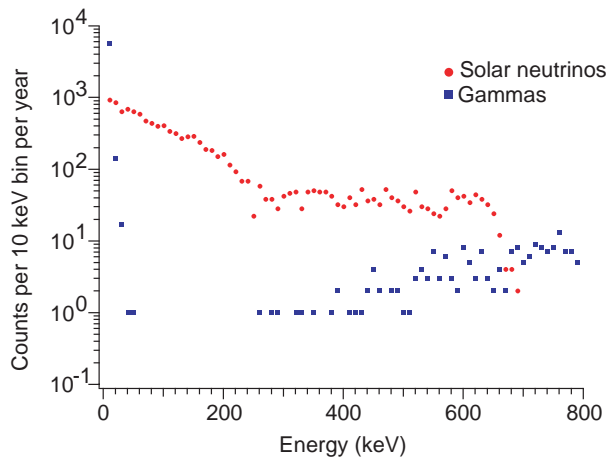, width=8cm} 
\caption{Simulated neutrino and
background rates in a CLEAN detector 6 meters in diameter, histogrammed
versus energy.  One year of simulation data are shown.  Results
correspond to a fiducial volume of $r$ $\leq$ 125 cm, and the data are
allocated into 10 keV bins.}
\label{fig:energy_histogram}
\end{center}
\end{figure}

Simulations have been performed for possible future CLEAN detectors of
various sizes.  The photomultipliers assumed for the simulations are
multichannel plate photomultipliers, with assumed U and Th levels of 3
ng/g.\footnote{If it is determined that the optimal photomultipliers
for CLEAN have somewhat higher levels of radioactivity, this will
necessitate having a slightly larger liquid neon volume to achieve a
given background level.  However, as photomultipliers used in current
neutrino experiments have U/Th levels on the order of 10-30 ng/g, and
since the high-density liquid neon is an effective gamma attenuator,
we expect the background levels indicated in Figures
\ref{fig:background} and \ref{fig:energy_histogram} to be reasonably
accurate predictions.} All gamma and x-rays from U and Th with
intensity greater than 1 \% were included.  Figure
\ref{fig:background} shows simulation results for several diameters of
possible future CLEAN detectors.  A fiducial volume corresponding to
Ê$r$ $\leq$ 12.6 cm corresponds to a neon fiducial mass 10 kg.  The
backgrounds arise from two sources: gamma rays that penetrate the
shielding layer to deposit energy within the fiducial volume, and
x-rays and low energy gammas that do not penetrate far and deposit
only a small amount of energy.  In the first case, there is little
background at low energies, since low-energy gammas are very unlikely
to penetrate the shielding layer.  In the second case, there is little
background at high energies, since high-energy events result in many
scintillation photons, and the positions of these events can be
accurately estimated.  Simulation results for a full-size CLEAN
detector have also been completed; the results are shown in Section
2.1, Figure \ref{fig:energy_histogram} and compared with the expected
solar neutrino signal.

The event reconstruction algorithm used in this study assumes that all 
detected photons produced by a particular event can be distinguished 
according to their arrival times from photon counts produced by other 
events. ÊThis is the only temporal information that we assume 
for event reconstruction. ÊIn contrast, in Super-K, SNO, KamLAND, 
and Borexino, event reconstruction algorithms are based on photon 
time-of-flight information. ÊIn the future, we plan to investigate the 
possibility of using additional photon time-of-flight information 
to improve the accuracy of our event reconstruction algorithm. 

\subsection{Internal backgrounds}

Because neon has a low binding energy on a variety of surfaces, we
expect to be able to rid neon of all contaminants.  These may be
divided into two categories: \textit{EUV absorbers} (such as $\rm
H_{2}$, $\rm N_{2}$, and $\rm O_{2}$) that could decrease the amount
of detected EUV light, and \textit{radioactive impurities} (such as
Ar, Kr, and Rn) that could occasionally decay, creating unwanted
backgrounds.  $\rm ^{39}Ar$ is produced by muon spallation in the
atmosphere and decays by $\rm \beta^{-}$ emission with an endpoint of
565 keV. The ratio of $\rm ^{39}Ar$/Ar has been measured to be (8.1
$\pm$ 0.3) $\times$ $10^{-16}$\cite{Ar}, and the full-size CLEAN
requires a natural Ar level of less than 1 $\times$ $10^{-10}$ g/g. 
$\rm ^{85}Kr$ is released from nuclear facilities as a fission product
of $\rm ^{235}U$ and $\rm ^{239}Pu$.  The ratio of $\rm ^{85}Kr$/Kr is
roughly 1.5 $\times$ $10^{-11}$, and the full-size CLEAN requires a
natural Kr level of less than 5 $\times$ $10^{-16}$ g/g\cite{Kr} to
ensure that this background does not exceed 1\% of the expected $p-p$
neutrino signal.

By passing the neon through cooled charcoal, all contaminants should
be removed with high efficiency.  The residence time for any species
on a surface is proportional to $\textrm{exp}(-E_{b}/kT)$.  All of the
contaminants have higher binding energies to charcoal than Ne (see
Table \ref{tab:impurities}), and at liquid neon temperature, $E_{b} >>
kT$ for all possible contaminants of concern.  A simple model of neon
gas placed in contact with an activated charcoal-filled cold trap
predicts a factor of $10^{6}$ Kr reduction per stage.

\begin{table}[tbhp]
    \centering
    \caption{Binding energies $E_{b}$ of various atoms and molecules
	to charcoal surfaces.  The well depth $D$ is also listed, since it
	approximates $E_{b}$ and is useful when $E_{b}$ has not been
	measured.  ``NM'' denotes that no measurement is listed in
	reference \cite{Vid91}. } 
    \label{tab:impurities}
    \begin{tabular}{|c|c|c|c|c|c|}
        \hline
       Species & $E_{b}$ (meV) & $D$ (meV) & Activity & $ T_{1/2}$ & $Q$ (keV) \\  
	   Ne & 30.1 & 32.6 & none & - & - \\
	   $\rm H_{2}$ & 41.6 $\pm$ 0.3 & 51.7 $\pm$ 0.5 & none & - & - \\
	   $\rm O_{2}$ & NM & 101.7 & none & - & - \\
	   $\rm N_{2}$ & NM & 104 $\pm$ 3 & none & - & - \\
	   Ar & 99 $\pm$ 4 & 96 $\pm$ 2 & $\rm ^{39}Ar$ ($\beta^{-}$) & 269 y & 565 \\
	   Kr & 126 & 125 $\pm$ 5 & $ \rm ^{85}Kr$ ($\beta^{-}$) & 10.76 y & 687 \\
	   \hline
  \end{tabular}
  
\end{table}

Radon will not be an important background in CLEAN. As in the case of
Ar and Kr, Rn will be removed by passing the Ne through cold traps. 
Also, because the diffusion of Rn through solid materials decreases
markedly with temperature, Rn will not enter the central tank through
the walls.  Any Rn originally in the neon will decay away with a
half-life of 3.8 days and will be negligible within a month of running
the experiment.  In this connection, we note that the Borexino
collaboration has demonstrated that by use of charcoal traps nitrogen
gas can be purified to a Rn activity of less than 0.1 $\rm Bq\,m^{-3}$
in the gas\cite{Bor02}.  Assuming that one can achieve equally low
levels in neon (this is a conservative assumption because the binding
energy of neon is less than that of nitrogen), the rate of Rn
daughters decaying in CLEAN should be negligible.

\subsection{Cosmogenic activities}

The full-scale CLEAN experiment must be located in an underground
facility to avoid radioactive backgrounds caused by muon spallation. 
While CLEAN will be surrounded by a muon veto to reject any prompt
radioactivity, some isotopes will be created that decay with long
half-lives and cannot easily be correlated with the muon pulses that
created them.  Many of the created nuclei decay via positron emission,
and therefore deposit at least 1.022 MeV in the neon, well above the
energy window of interest.  Others, such as $\rm ^{14}C$ and $\rm
^{10}Be$, decay with very long half-lives and are unlikely to create
an appreciable background.  This leaves $\rm ^{23}Ne$, $\rm ^{21}F$,
$\rm ^{20}F$, $\rm ^{20}O$, $\rm ^{19}O$, $\rm ^{17}N$, $\rm ^{16}N$,
$\rm ^{15}C$, $\rm ^{11}Be$, $\rm ^{7}Be$, and $\rm ^{3}H$, all of
which could create backgrounds in the energy range of interest.  Most
of these are beta emitters with endpoint energies of several MeV, and
the majority of the decays will not create much background.  One
exception is $\rm ^{7}Be$; this decay has an endpoint of only 862 keV.
Another exception is $\rm ^{3}H$ with an endpoint of 18 keV; however
the energy deposited by this decay is rather small and most of these
events will be below the CLEAN threshold.

While the cross-sections for the creation of these isotopes have not
been measured, we can roughly estimate their production rates based on
careful studies of muon spallation cross-sections in $\rm ^{12}C$ and
other nuclei.  For example, the Borexino collaboration has measured a
$\rm ^{7}Be$ production cross-section of 230 $\pm$ 23 $\rm \mu barn$
on $\rm ^{12}C$ at a muon energy of 190 GeV\cite{Hag00}. 
Conservatively assuming a much higher muon-neon spallation
cross-section of 1 millibarn for the creation of each of the isotopes
of concern, and assuming a muon flux of 0.23 $\rm m^{-2}\,day^{-1}$
(the muon flux in SNOLAB), only 2.1 events per year would occur within
the 10,000 kg fiducial volume and within the 0 - 800 keV range of
interest.  If CLEAN were located at Gran Sasso, with a higher muon
flux of 8.2 $\rm m^{-2}\,day^{-1}$, then about 100 events per year
would be expected within the energy window of interest.

Neutron absorption on Ne is not expected to contribute significant
backgrounds.  Both fast and slow neutrons, created largely from
$\mathrm{U}(\alpha,n)$ reactions in the rock, will be absorbed by
water shielding surrounding the liquid neon.  Some high energy
neutrons created by muon spallation may penetrate the water shielding,
but at a deep site like SNOLAB the flux of these high energy neutrons
is small.  If neutrons are captured in the neon, then the unstable
neon isotope $\rm ^{23}Ne$ (created by neutron absorption on $\rm
^{22}Ne$) can create decays in the energy range of interest.  However,
the neutron absorption cross-section of $\rm ^{22}Ne$ is small (46
mbarn for thermal neutrons), and the natural abundance of $\rm
^{22}Ne$ is only 9.2 \%.

\section{The Physics Potential of CLEAN}

The CLEAN concept has multiple applications in detection of rare
events.  Here we discuss its application to four of them: $p-p$
neutrino detection, the search for the neutrino magnetic moment, and
detection of supernova neutrinos.

\subsection{Solar neutrinos}

The study of neutrinos plays a prominent role in astrophysics and
particle physics.  Though they are emitted in vast numbers by stars
and can be easily made in modern particle accelerators, neutrinos are
difficult to detect because they have no charge and interact only
through the weak force.  Though the neutrino was first hypothesized in
1930 by Pauli to explain the observed characteristics of beta
decay\cite{Pau30}, it was many years later (1953) before the existence
of the neutrino was confirmed experimentally by Reines and Cowan\cite{Rei53}.

The dominant source of neutrinos is the Sun.  While 97.7 \% of
the Sun's energy is emitted as electromagnetic radiation, another
2.3 \% is emitted in the form of neutrinos\cite{Bah89}.  And while the
light takes about 10,000 years to diffuse out of the Sun, the
neutrinos leave immediately.  These neutrinos originate from a variety
of nuclear reactions in the solar core known as the ``$p-p$
chain".  

In order to verify that the Sun is powered by nuclear fusion
reactions, the first solar neutrino experiment (based on neutrino
absorption by Cl) was built in 1968 by Davis and
collaborators\cite{Dav68}.  While this experiment proved that the Sun
is powered by nuclear reactions, the number of detected neutrinos fell
well short of astrophysical predictions.  In the years since this
pioneering experiment, other detectors verified this shortfall.  Two
detectors based on neutrino absorption in Gallium,
GALLEX\cite{Ans92,Ham99} and SAGE\cite{Aba91,Abd99}, showed that the
effect extended to low energy neutrinos, and the
Kamiokande\cite{Hir90} and Super-Kamiokande\cite{Fuk99a,Fuk99b}
detectors showed that the dearth of detected neutrinos extends across
the entire solar neutrino spectrum.  In addition, the sophistication
of models of the solar interior have advanced, resulting in a sharper
discrepancy with theory\cite{Bah98}.  Therefore, recent years have
seen a focusing of attention on the fundamental properties of the
neutrino itself, since models that include neutrino mass and flavor
mixing could cause the $\rm \nu_{e}$ emitted by the sun to be
converted into $\rm \nu_{\mu}$ and $\rm \nu_{\tau}$, which cannot be
absorbed by nuclei and have smaller cross-sections for
neutrino-electron scattering.

The leading model for this flavor conversion is that neutrinos have
mass, and that the neutrino mass matrix is non-diagonal when expressed
in the flavor basis.  The weak eigenstates $|\mathrm{\nu}_{a}\rangle$
($a$ = e, $\rm \mu$, $\rm \tau$) will be in this case linear
superpositions of the mass eigenstates $|\mathrm{\nu}_{i}\rangle$ ($i$
= 1,2,3), where $|\mathrm{\nu}_{a}\rangle\, =
\,\sum_{i}U_{a,i}|\mathrm{\nu}_{i}\rangle.$ If neutrinos have mass and
mix according to this equation, then this results in neutrino
oscillations; a neutrino that was initially in the weak eigenstate $a$
can be transformed into another weak eigenstate $a'$.  This
oscillation has been observed for muon neutrinos, which change flavor
after being created in the upper atmosphere (as shown by the
Super-Kamiokande experiment\cite{Fuk98}).  Most recently, new results
from SNO have proven decisively that solar neutrinos originally of the
electron flavor are largely converted to $\rm \mu$ or $\rm \tau$
neutrinos by the time they reach the Earth\cite{SNO,Ahm03}.  The
mechanism by which this oscillation occurs is believed to be the
Mikheyev-Smirnov-Wolfenstein (MSW) effect, in which the neutrino
states are mixed by matter interactions as they pass out of the Sun. 
In a simple 2 $\times$ 2 mixing scenario (approximately valid as long
as $U_{e3}$ is sufficiently small), one finds that the oscillation
probability depends only on the difference $\Delta
m^{2}=m_{i}^{2}-m_{j}^{2}$ between the squares of the mass eigenstates
and one mixing angle $\theta$.  The most heavily favored scenario is
the MSW solution known as Large Mixing Angle, or LMA for short.  The
LMA solution has recently been confirmed by KamLAND\cite{Egu03}, an
experiment testing for oscillation of reactor antineutrinos.

Three compelling arguments yield the conclusion that the next large
solar neutrino experiment should be designed to detect low energy
solar neutrinos from the $p-p$ reaction, $p + p \rightarrow d + e^{+} + \nu_{e}.$

First, the $p-p$ neutrinos account for about 91 \% of the solar
neutrino production according to the standard solar model (SSM).  The
resulting predicted flux at the Earth ($5.94 \times 10^{10}\, \rm
cm^{-2}\, s^{-1}$) can be calculated to an uncertainty of
1 \%.\cite{Bah98} As such, the $p-p$ flux is the most precisely
predicted by the SSM. This spectacularly precise prediction calls out
for an experiment to test it.  Such a measurement would constitute an
important test of the mechanism by which stars shine, with significant
implications for all of astrophysics.  An accurate $p-p$ neutrino
experiment will also likely be able to measure the $\rm ^{7}Be$
neutrino flux.  A measurement of the ratio of the $p-p$ flux to the
$\rm ^{7}Be$ flux can determine the relative rates of
the two termination reactions of the $p-p$ chain in the solar
interior, $\rm ^{3}He-^{3}He$ and $\rm ^{3}He-^{4}He$, thus 
testing this prediction of the standard solar model\cite{Bah03}.

Second, if the correct solution is LMA, then the neutrino $\rm \Delta
m^{2}$ and $\rm tan^{2}\theta$ parameters can be best determined with
a combination of experiments.  If the SSM is determined to be
accurate, then a high-statistics $p-p$ neutrino experiment would yield
a particularly good measurement of the mixing angle.  This would
nicely complement KamLAND, which will give an accurate measurement of
$\rm \Delta m^{2}$. A measurement of the $p-p$ flux has an energy 
range very
different from that of the $\rm ^{8}B$ experiments, and uses
neutrinos rather than the antineutrinos of KamLAND. It is also
interesting to note that the LMA solution behaves qualitatively
differently at low neutrino energies, where neutrino flavor conversion
is dominated by vacuum oscillations, rather than by matter-induced
oscillations as in the $\rm ^{8}B$ neutrinos detected by
Super-Kamiokande and SNO. Thanks to these differences, and thanks to
the precise prediction of the $p-p$ neutrino flux, an accurate
measurement of the $p-p$ flux would provide an independent test of the
robustness of the neutrino mixing picture that is being assembled.

Third, a combination of low-energy experiments measuring different
combinations of charged-current (CC) and neutral-current (NC)
reactions would yield the total solar neutrino flux independent of
flavor.  By comparing this measured total flux to the predicted
``standard candle'' flux, tight limits can be placed on a possible
sterile neutrino component.  While SNO has shown that sterile
neutrinos alone do not explain the solar neutrino problem, it
nevertheless does not rule out a sizable sterile component\cite{SNO}. 
The current $\rm 1\sigma$ limit on an active-sterile admixture is
$\rm \sin^{2}\eta \leq 0.16$, and the best chance for a significant
improvement in sensitivity to sterile neutrinos is through the
accurate measurement of the $p-p$ neutrino flux\cite{Bah03}.

It is also likely that a new technology capable of a $p-p$ neutrino
measurement could also measure the CNO neutrino flux through analysis
of the neutrino spectrum in the energy window of 0.7 to 1.0 MeV (above
the $\rm ^{7}Be$ neutrino scattering spectrum), provided that the
detector is built to a sufficiently large scale.  We have recently estimated
the statistical uncertainty for such a measurement through Monte Carlo
simulation\cite{Ros03} and conclude that for a CLEAN detector of
9 m diameter and 100 metric ton fiducial mass, the CNO flux could be measured
with a statistical uncertainty of 13 \% with one year of data taking,
assuming that the detector is deep enough that muon-induced
backgrounds are not significant.  There is currently no measurement of
the CNO flux; the experimental upper limit on this flux is estimated
to be 7.3 \% of the total solar neutrino flux\cite{Bah02}.

We have investigated the ability of CLEAN to measure the p-p neutrinos
through simulations of radioactive backgrounds, assuming that these
backgrounds are dominated by gamma ray emission from the phototubes. 
Preliminary simulations have been performed for the full-size CLEAN,
showing that for a neon tank of 6 meter diameter (135-ton total
neon mass), these backgrounds are significantly smaller than the
expected neutrino signal (about 2900 events per year assuming the LMA
solution) in the energy range of interest (see Figure
\ref{fig:energy_histogram}).  As in the simulations of smaller
detectors, the backgrounds at low energies are determined by position
resolution, while the backgrounds at higher energies are dominated by
gamma rays that manage to penetrate the shielding layer.  Based on
these simulations, a neutrino detection threshold of around 20 keV or
less is likely.

\subsection{Neutrino magnetic moment}

As well as having masses, neutrinos could have magnetic moments. 
The best experimental limits on the magnetic moment of the neutrino
come from experiments that look for deviations of the cross-sections
for $\nu - e^{-}$ and $\bar{\nu} - e^{-}$ scattering from those expected
from the weak interaction alone.  If the electron neutrino has a
magnetic moment $\mu_{e}$, then it can scatter from electrons by
magnetic dipole-dipole scattering, enhancing the scattering
cross-section at low recoil energies.  An additional term is then
added to the scattering cross-section: $$ (\frac{d\sigma}{dT})_{\mu} =
\frac{\pi\alpha^{2}\mu_{e}^{2}}{m_{e}^{2}}\frac{1-T/E_{\nu}}{T}.$$
In this expression, $E_{\nu}$ in the initial neutrino energy and $T$
is the electron recoil energy.  From this expression, it can be seen 
that if the neutrino has a magnetic moment, then the neutrino-electron 
scattering cross-section grows as 1/T at low recoil energies. 
Sensitivity to the neutrino magnetic moment can be enhanced by looking 
for neutrino-electron scattering events at low energy, where the 
electromagnetic scattering term is large. For this reason, it helps to have a
low energy threshold when searching for a neutrino magnetic moment.

The lowest limits on the neutrino magnetic moments come from both
solar neutrino experiments (which set limits on $\mu_{\nu}$) and
reactor experiments (which set limits on $\mu_{\bar{\nu}}$).  In the
former category, data from Super-Kamiokande\cite{Fuk99b} set the tightest
limit to date, $\mu_{\nu} < 1.5 \times 10^{-10} \mu_{B}$ \cite{Bea99}.  In the
reactor experiments, a limit of $1.3 \times 10^{-10} \mu_{B}$ (90 \%)
was set by the TEXONO collaboration\cite{Won03} using an
ultra-low-background germanium counter, while the MUNU
collaboration\cite{Dar03} has set a limit of $1.0 \times 10^{-10}
\mu_{B}$ (90 \%) using a $\rm CF_{4}$-filled time-projection chamber. 
These two experiments had substantially different approaches.  In the
TEXONO experiment, a very low threshold was achieved, but internal
backgrounds ultimately limited its sensitivity.  In the MUNU
experiment, very low backgrounds were achieved through a sophisticated
gamma veto system and by comparing forward versus backward scattering
rates.  However, the threshold was much higher, about 600 keV.

Using a CLEAN detector, new limits could be placed on both $\mu_{\nu}$
and $\mu_{\bar{\nu}}$. First, a small (1 metric ton) liquid neon filled
detector, placed near a nuclear reactor, would be very sensitive to
neutrino-electron scattering, with a combination of low internal
backgrounds and low threshold.  With a 1-tonne detector, we estimate
that the limit on $\mu_{\bar{\nu}}$ could be decreased by
approximately two orders of magnitude.  This experiment would have the
additional bonus of measuring the low-energy neutrino flux from a
reactor, something that has never previously been achieved.  Second,
by measuring the $p-p$ solar neutrino spectrum and comparing its
spectral shape to that expected from the SSM and MSW oscillations, one
may put a limit on the neutrino magnetic moment.  Preliminary analysis
shows that the full-size CLEAN, with a low energy threshold and low
background rate, would be sensitive to a neutrino magnetic moment as
low as 1 $\times$ $10^{-11}$ $\rm \mu_{B}$.

\subsection{Supernova neutrinos}

Following the gravitational collapse of a star with mass greater than
eight solar masses, a type II supernova releases an energy of about $3
\times 10^{53}$ ergs in just a few seconds.  Though supernovae can be
seen visually, only 1 \% of the energy is emitted as light.  The
remainder is carried away by electron, mu, and tau neutrinos and
antineutrinos, creating a large mixed neutrino flux at the Earth. 
Existing neutrino detectors such as Super-Kamiokande and SNO would
easily detect these supernovae, just as SN1987A was detected by
Kamiokande and IMB. However, these detectors do not yield much
neutrino spectral information, and Super-Kamiokande is sensitive largely
to electron anti-neutrinos and would give little information
about the more energetic $\rm \nu_{\mu}$ and $\rm \nu_{\tau}$.

 In a recent paper, we point out that the $\rm \nu_{\mu}$ and $\rm
 \nu_{\tau}$ neutrinos could be readily detected using a
 ``flavor-blind'' detector sensitive to neutrino-nuclear
 scattering\cite{Hor02}, provided that it had a low energy threshold. 
 With high sensitivity to neutral current interactions, such a
 low energy neutrino detector yields additional information about the
 dynamics of supernova collapse, the flavors of neutrinos emitted
 during a supernova, and the energy spectra of these neutrinos. 
 Measurement of the $\nu_{\mu}$ and $\nu_{\tau}$ fluxes and energies
 would also give an accurate determination of the total energy
 released by the supernova.  A similar approach is described in detail
 by Beacom \textit{et al.}, who discuss the possible detection of
 neutrino-proton scattering at KamLAND and Borexino\cite{Bea02}.
 
Another approach to detecting heavy supernova neutrinos is through
inelastic reactions on oxygen or carbon
nuclei\cite{Woo90,Pie01,Cad02}.  However, these reactions have
relatively low yield, and neutrino energy information is lost since
the outgoing neutrino is not detected.

The nuclear elastic scattering approach is promising because the coherent
elastic scattering cross-section is large, and because all neutrino
components contribute to the signal, allowing one to learn about
$\nu_{\mu}$ and $\nu_{\tau}$ fluxes from the supernova.  
Assuming equal partitioning
in energy among the $\nu_{e}$, $\bar{\nu}_{e}$, $\nu_{\mu}$,
$\bar{\nu}_{\mu}$, $\nu_{\tau}$, and $\bar{\nu}_{\tau}$ components,
we find that a standard supernova at 10
kiloparsecs gives a total yield of 3.99 events per metric ton in liquid Ne,
with most of these (3.08) coming from $\nu_{\mu}$, $\bar{\nu}_{\mu}$,
$\nu_{\tau}$, and $\bar{\nu}_{\tau}$.  Because the supernova events
occur within 10 seconds, a large fiducial volume could be used with 
very little background.  For a
100 tonne fiducial volume, a total of 400 $\nu$-Ne elastic events (285
above a 10 keV threshold) could be detected from a supernova 10
kiloparsecs away.

The practical threshold attainable in CLEAN for supernova neutrinos
depends on the quenching factor for nuclear recoils in liquid Ne. 
However, this factor should not be too small, as the denser liquid Xe
has a quenching factor of 0.22 $\pm$ 0.01\cite{Aki02}.  Though the
quenching factor in liquid Ne has not yet been measured, one expects
less quenching than the denser liquid Xe.  Should a supernova occur,
CLEAN could detect nearly all events above threshold, which could be
as low as 4 keV.  

\section{Ongoing research and development}

Before building a full-size CLEAN experiment, it makes sense to better
study the scintillation and optical properties of liquid neon, test
the scintillation response of liquid neon for nuclear recoils, build
and test neon purification equipment, and test photomultipliers at
low temperature.  This is all part of the ongoing CLEAN research
program.

One of the first projects is to build a 30 kg detector (micro-CLEAN). 
The liquid neon could be surrounded by as many as 10 photomultipliers,
and much of the tank volume could be filled by acrylic or copper to
reduce the total amount of neon to be liquified for a given test.  A
photomultiplier support structure will be built to allow various
detector geometries.  Using this apparatus, we intend to test:
\begin{itemize}
	\item The light yield of liquid neon.
	\item Absorption and scattering of neon scintillation light.
	\item The coupling of the scintillation cell to neon purification systems.
	\item The performance of photomultipliers immersed in liquid neon.
	\item Methods of calibrating the detector for different radiation 
	types.
	\item Pulse shape analysis for different radiation types.
	\item Development of calibrated event reconstruction algorithms for test geometry 
\end{itemize}

Tests of scattering and absorption would be performed using collimated
scintillation sources and photomultipliers in a carefully chosen
geometry. The best geometry for these tests would be determined 
through Monte Carlo simulation. Based on previous measurements of 
scattering and absorption in organic scintillator done at the Borexino 
Counting Test Facility\cite{Joh95}, we believe that a spherical 
geometry will be necessary to disentangle the effects of scattering 
and absorption.  By placing a radioactive source inside a tube with one
open end, while allowing the liquid neon to enter the tube, a
collimated scintillation source can be created. Comparing 
the scintillation light detected by a photomultiplier facing the 
source to that detected by a photomultiplier that can detect only the 
scattered light, the scintillation scattering length can be 
measured. By comparing these measurements for different distances from 
the source to the photomultipliers, the absorption and scattering 
lengths can be determined independently. 

A key component of CLEAN is the purification of liquid neon, removing
elements that could absorb the EUV scintillation light or contribute
to radioactive background.  We expect to be able to rid neon of these
contaminants by passing it through an adsorbant such as activated
charcoal.  In a working CLEAN experiment, two separate purification
systems would be needed.  The first purification system will use
getters and liquid nitrogen-cooled charcoal traps to purify gaseous
neon before liquification.  The second purification system will be
used to continuously purify liquified neon, removing liquid neon from
the scintillation cell and returning it immediately, while allowing
the liquid neon scintillation properties to be continuously monitored
during purification.  The apparatus would be housed in a bucket dewar
and be surrounded by a liquid nitrogen bath.  The system would use a
cryorefrigerator to pull the liquid into a small chamber, then
pressurize the liquid neon through a cold trap by closing the entrance
valve and warming the neon.  This approach has few moving parts and
less opportunity for vacuum leaks to occur than if a mechanical pump
were used to circulate the liquid neon.  The system would make use of
several cryogenic valves, which are available commercially.  Two
alternating cold traps would be used; while one purifies neon, the
other is warmed and pumped out.  The two neon purification systems
have quite different technical requirements.  The first system will be
optimized for cleaning efficiency, but need not have high throughput,
as the neon liquification rate will be limited by cooling power.  The
continous purification system will be optimized for neon flow rate,
but need not have high purification efficiency.  As the neon will be
removed and replaced continuously, the best purification efficiency
one can achieve is 1/$e$ per volume change.

We also intend to test a variety of photomultipliers at 27 K and to
measure their radioactivity levels in order to determine the best
light detector for use in CLEAN. Most photomultipliers used in nuclear
and particle physics research are made with glass, which contains
significant levels of radioactive U, Th, and K impurities. 
State-of-the-art photomultipliers with ``ultralow background glass''
have U and Th levels of about 30 ng/g, and a K level of about 60
ng/g\cite{ET,disclaimer}.  However, it is quite possible that glass-less
photomultipliers, with a mechanical structure composed of fused
silica, metal, and sapphire, could have U, Th, and K levels on the
order of 1 ppb.  Though we would like to develop new photomultipliers
that have low radioactive impurities and function at low temperature,
photomultipliers that are currently on the market could be used as the
photon detectors in CLEAN. Photomultipliers are currently being used
by the ICARUS collaboration to detect scintillations in liquid argon
(90 K), and we expect these to function at 27 K as well.

To learn how sensitive CLEAN would be to nuclear scattering events,
such as those arising from neutrino-nuclear scattering, the
liquid neon scintillation should be calibrated with a neutron source. 
This procedure is common, especially for dark matter searches, and the
approach in CLEAN would be similar to that followed in recent
calibrations of liquid xenon-based detectors\cite{Aki02,Ber01}.  In
order to perform this calibration, we would purchase a commercial
neutron source that produces 2.5 MeV neutrons via the $d + d
\rightarrow n + \mathrm{^{3}He}$ reaction.  We would detect scattered
neutrons in coincidence with the flash of scintillation light from the
liquid neon.  By knowing the scattering angle $\theta$, the neutron
energy, and amount of light detected, the experiment may be
calibrated for nuclear recoils. 

Excitation of the liquid neon with neutrons will also allow us to
study the scintillation pulse shape from nuclear recoils in liquid
neon.  Pulse shape analysis has proven to be useful in experiments
using liquid xenon as a target for WIMP detection\cite{Aki02,Ber01},
and we have previously studied liquid helium and found that it
exhibits significantly different scintillation pulses for alpha and
beta excitation\cite{McK03a,McK03b}.  A better understanding of how
pulse shape varies with excitation density may allow us to also
discriminate between radiation types in liquid neon.  Given the very
different times scales of triplet and singlet scintillation in liquid
neon, it is likely that pulse shape discrimination in liquid neon will
be quite effective.

\section{Conclusion}

The CLEAN approach to low background detection of rare events has
potential application to low energy neutrino detection, supernova
neutrino detection, and the search for the magnetic dipole moment of the
neutrino.  While the basic technique
(use of a wavelength shifter and phototubes in combination with a
liquified noble gas as detection medium) could in principle be used
with liquid helium or xenon, liquid neon appears for the time being to
be the most promising option.  Ongoing research promises to determine
liquid neon's light yield and pulse shape for a variety of radiation
types, measure the optical properties of liquid neon for its own
scintillation light, test photomultipliers immersed in liquid neon,
and develop techniques for neon purification.  There also remains work
to be done on the event reconstruction algorithm, including empirical
calibration of the event reconstruction algorithm, developing models
for the transition matrix, possible incorporation of time-of-flight
information, accounting for possible absorption effects, and
development of approaches for cylindrical geometries.

\section{Acknowledgements}

The authors thank A. Hime, E. Kearns, and T. Shutt for
valuable discussions.  The work of D. M. is supported by NSF grant
PHY-0226142.  Contributions of NIST (an agency of the US Government)
to this work are not subject to copyright.


\begin{thebibliography}{99}
\bibitem{disclaimer} Products or companies named here are cited only in the interest of 
complete description, and neither constitute nor imply endorsement by 
NIST or by the US government. Other products may be found to serve 
just as well.
\end{thebibliography}
\end{document}